\documentclass[aps,prb,twocolumn,amsmath,longbibliography]{revtex4}  
\usepackage{bbm}
\usepackage{mathrsfs}
\usepackage{amsmath}
\usepackage{amsfonts}
\usepackage[colorlinks=true,citecolor=blue,anchorcolor=blue]{hyperref}
\usepackage{graphicx,epstopdf}
\usepackage{subfigure}
\usepackage{epsfig}
\usepackage{dcolumn}
\usepackage{bm}
\usepackage{color}
\usepackage{natbib}
\usepackage{amssymb}
\usepackage{xcolor}
\usepackage{braket}

\begin{document}

\title{Floquet engineering of magnetic topological insulator MnBi$_2$Te$_4$ films}

\author{Tongshuai Zhu$^1$, Huaiqiang Wang$^1$ and Haijun Zhang$^{1,2,*}$}

\affiliation{
 $^1$ National Laboratory of Solid State Microstructures, School of Physics, Nanjing University, Nanjing 210093, China\\
$^2$ Collaborative Innovation Center of Advanced Microstructures, Nanjing University, Nanjing 210093, China\\
}

\begin{abstract}
Floquet engineering is an important way to manipulate the electronic states of condensed matter physics. Recently, the discovery of the magnetic topological insulator MnBi$_2$Te$_4$ and its family provided a valuable platform to study magnetic topological phenomena, such as, the quantum anomalous Hall effect, the axion insulator state and the topological magnetoelectric effect. In this work, based on the effective model and first-principles calculations in combination with the Floquet theory, we reveal that the circularly polarized light (CPL) induces the sign reversal of the Chern number of odd-septuple-layer (SL) MnBi$_2$Te$_4$ thin films. In contrast, the CPL drives the axion insulator state into the quantum anomalous Hall state in even-SL MnBi$_2$Te$_4$ thin films. More interestingly, if the topmost van der Waals gap between the surface layer and the below bulk in MnBi$_2$Te$_4$ films is slightly expanded, a high Chern number $|C|=2$ can be realized under the CPL. Our work demonstrates that the light can induce rich magnetic topological phases in MnBi$_2$Te$_4$ films, which might have potential applications in optoelectronic devices.
\end{abstract}

\email{zhanghj@nju.edu.cn}

\maketitle

\section{Introduction}
Topological insulators with gapless surface states have become an important concept in condensed matter physics over the past decade\cite{Qi2011Topological,Hasan2010Colloquium}. When magnetism is induced in topological insulators, rich magnetic topological states would emerge, such as, the quantum anomalous Hall (QAH) state and the axion insulator state\cite{Haldane1988,qi2008,li2010}. As a typical magnetic topological state, the QAH state was predicted and realized in Cr-doped (Bi,Sb)$_2$Te$_3$ films\cite{yu2010,chang2013experimental}, where the magnetism is not intrinsic but induced through the doping process. Interestingly, MnBi$_2$Te$_4$ and its family were recently discovered to be intrinsic magnetic topological insulators\cite{gong2019experimental,zhang2019topological,li2019intrinsic,otrokov2019prediction,chen2019topological,hao2019gapless,li2019gapless,rienks2019large,chen2019intrinsic,vidal2019topological,klimovskikh2020tunable,zhu2021tunable,wang2020heterostructures,zhang2020MBT225,Yang2022Evolution,Zhan2021A}. MnBi$_2$Te$_4$ consists of septuple layers (SL) with an A-type antiferromagnetic (AFM) ground state, and the SLs are coupled together by the weak van der Waals (vdW) interactions\cite{zhang2019topological}. A QAH state was observed experimentally in the odd-SL MnBi$_2$Te$_4$ films\cite{deng2020quantum}, and an axion insulator state with a zero Hall plateau was also observed in even-SL MnBi$_2$Te$_4$ films\cite{liu2020robust}. The control and engineering of magnetic topological states in MnBi$_2$Te$_4$ and its family with the intrinsic magnetism have become one of the most important topics in the field.

In order to engineer the properties of materials, light irradiation provides a powerful method, which does not need to be in direct contact with the sample in experiments. Recently, Floquet engineering via periodic light opens a route toward engineering exotic Floquet topological states with high tunability. Many Floquet topological states have been theoretically proposed or experimentally realized in optically driven systems, such as,  the Floquet topological insulator in semiconductor quantum wells\cite{lindner2011floquet}, the gapped surface Dirac cone in a topological insulator\cite{Wang2013Observation}, light-induced QAH effect\cite{mciver2020light,Kitagawa2011Transport,xu2021light,kong2022floquet,ning2022photoinduced,wang2018Light,Yap2017Computational}, the Floquet engineering of magnetism in topological insulator thin films\cite{liu2021floquet,qin2022phase}, topological phase transitions in semimetals\cite{Yan2016Tunable,hubener2017creating,Liu2019Engineering,Zhang2018Engineering,Li2018Realistic,Zhu2021Floquet,Zhou2016Floquet,Liu2018Photoinduced}, and so on \cite{Gomez-Leon2013Floquet,Rudner2013Anomalous,Ezawa2013Photoinduced,Grushin2014Floquet,mahmood2016selective,Oka2019Floquet,Mikami2016Brillouin,Eckardt2015High,Ma2021Floquet,Pervishko2018Impact, Cheng2019, Wang2017, Bomantara2016, Bomantara2018}.

In this work, based on the effective model and first-principles calculations in combination with the Floquet theory, we investigate topological phase transitions in MnBi$_2$Te$_4$ films irradiated by a circularly polarized light (CPL). For the odd-SL MnBi$_2$Te$_4$ thin films, a right circularly polarized light (RCPL) can drive the QAH state into a normal insulator (NI) state and then to another QAH state with a sign reverse of the Chern number, while a left circularly polarized light (LCPL) can greatly enhance the band gap as the amplitude of the light increases, which indicates that the critical temperature of QAH states can be improved by an LCPL. In contrast, we find that the CPL can also tune the axion insulator state into a QAH state in the even-SL MnBi$_2$Te$_4$ thin films. In addition, when the topmost vdW gap on the surface is slightly expanded, possibly by the experimental process or impurities\cite{Eremeev2012effect, wang2022three}, the energy gap of surface Dirac-cone states of MnBi$_2$Te$_4$ can be greatly decreased and even become gapless. Here, we reveal that a higher QAH plateau with the Chern number $|C|=2$ can be achieved under a CPL for MnBi$_2$Te$_4$ films with a slight expansion of the topmost vdW gap.


\section{Method}
\label{method}

\begin{figure}[t]
  \centering
  \includegraphics[width=3.4in]{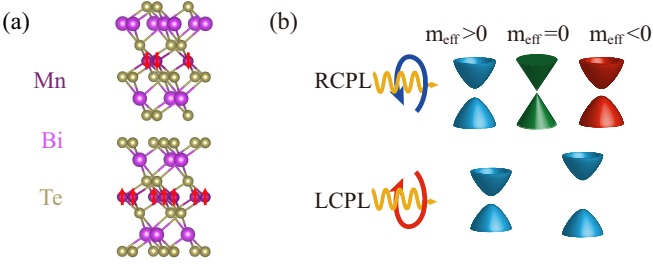}\
  \caption{(a) The crystal structure of MnBi$_2$Te$_4$. The red arrow denotes the spin moment of the Mn atom. (b) Illustration of the effect of the CPL on the Dirac-cone state. The RCPL leads to the band gap closing and reopening processes and the LCPL  enhances the band gap. }\label{fig1}
\end{figure}

 \subsection{The effective model with Floquet theory}

\begin{figure}[t]
  \centering
  \includegraphics[width=3.4in]{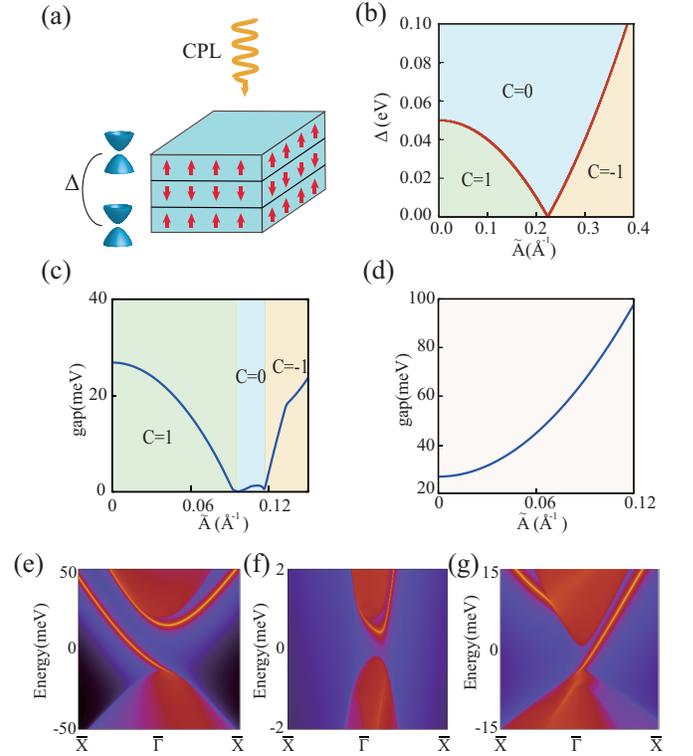}\
  \caption{The light engineering on the odd-SL MnBi$_2$Te$_4$ thin films. (a) The schematic of the 3-SL MnBi$_2$Te$_4$ under a CPL. Two massive (gapped) Dirac-cone surface states are on the top and bottom surfaces and the coupling strength of the two Dirac cones is $\Delta$. (b) RCPL-induced topological phase diagram calculated by the effective model for odd-SL MnBi$_2$Te$_4$ thin films, the Zeeman coupling strength is $m=50$ meV. (c,d) The band gap of 3-SL MnBi$_2$Te$_4$ from the first-principles calculations for RCPL (LCPL) vs $\tilde{A}$. The edge state of 3-SL MnBi$_2$Te$_4$ film under different strength of the light for (e) $\tilde{A}$ = 0.06 \AA$^{-1}$, (f) $\tilde{A}$ = 0.11 \AA$^{-1}$  (g) $\tilde{A}$ = 0.12 \AA$^{-1}$. The frequency of the CPL is $\omega=1$ eV ($\sim2.42\times 10^{14}$ Hz).   }\label{fig2}
\end{figure}

\begin{figure}[t]
  \centering
  \includegraphics[width=3.4in]{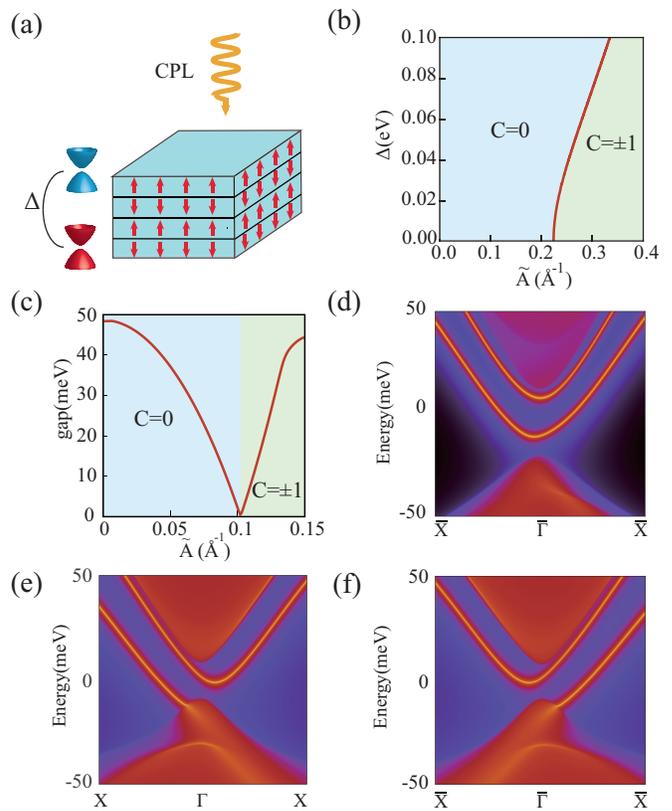}\
  \caption{The light engineering on the even-SL MnBi$_2$Te$_4$ thin films. (a) The schematic of the 4-SLs MnBi$_2$Te$_4$ under a CPL. Massive Dirac-cone surface states are on the top and bottom surfaces and the coupling strength of the two Dirac cones is $\Delta$. The blue (red) Dirac cone represents the sign of the mass term $m>0$ ($m<0$). (b) CPL-induced topological phase diagram for even-SL MnBi$_2$Te$_4$ thin films calculated by the effective model, the Zeeman coupling strength is $m=50$ meV. (c) The band gap of 4-SL MnBi$_2$Te$_4$ varies with $\tilde{A}$ from the first-principles calculations for the CPL.  The edge state of  4-SLs MnBi$_2$Te$_4$ under the RCPL of $\tilde{A}$ = 0.06 \AA$^{-1}$ (d), the LCPL of $\tilde{A}$ = 0.12 \AA$^{-1}$. (e)the RCPL of $\tilde{A}$ = 0.12 \AA$^{-1}$. The frequency of the CPL is $\omega=1$ eV. }\label{fig3}
\end{figure}

For MnBi$_2$Te$_4$ thin films, schematically shown in Figs.~\ref{fig2}(a) and~\ref{fig3}(a),  the surface states can be described by a low-energy effective model composed of two Dirac cones from the top and bottom surfaces,

 \begin{equation}
 \label{eq1}
\begin{split}
H(\bm{k})=\begin{pmatrix}
h_1(\bm{k})&\Delta I_{2}\\
\Delta I_{2}&h_2(\bm{k})
\end{pmatrix}
\end{split}
\end{equation}
where the Hamiltonian of each  Dirac cone is described as $h_i(\bm{k})=s_iv(k_x\sigma_y-k_y\sigma_x)+m_i\sigma_z$, with $i=1$ (or 2) denoting the top (or bottom) surface, and $s_i=\pm1$ representing the helicity of the surface Dirac cones. $v$ is the Fermi velocity, which is set as $v=1$ for simplicity in our model analysis. $m_i=\pm m$ indicates the Zeeman coupling induced by the time-reversal-symmetry-breaking magnetic moment on each surface. For odd-SL MnBi$_2$Te$_4$ thin films,  we have $m_1=m_2=m$, while for even-SL MnBi$_2$Te$_4$ thin films, we have $m_1=-m_2=m$. $\Delta$ describes the coupling strength between the Dirac cones on the top and bottom surfaces due to the finite-size effect. $I_{2}$ is the $2\times 2$ identity matrix.

 When a CPL is applied, the time-dependent Hamiltonian is obtained by making Peierls substitution $\bm{k}\rightarrow\bm{k}+e\bm{A}/\hbar$, where $\bm{A}$ is the vector potential of the light. For a CPL incident along the [001] direction of the sample, the vector potential $\bm{A}$ is written as $\bm{A}=A_0(\cos\omega t,\eta\sin\omega t,0)$, where $\eta=+1$ ($-1$) denotes the RCPL (LCPL),  $\omega$ is the frequency of the CPL, and $A_0$ is the amplitude of the vector potential. We define $\tilde{A}={eA_0}/{\hbar}$.  In this paper, we consider the off-resonant case with $\omega$ significantly larger than the system's typical energy scale ($\omega\gg m, \Delta$), where one can apply the off-resonant approximation and obtain an effective modified static Hamiltonian\cite{Yan2016Tunable,Kitagawa2011Transport}

\begin{equation}
\begin{split}
H_{\mbox{\tiny eff}}(\bm{k})=H_0(\bm{k})+\sum_{n\ge 1}\frac{[H_{-n},H_n]}{n\omega}+O(\frac{1}{\omega^2}),
\end{split}
\end{equation}
where $H_n=\frac{1}{T}\int^T_0H(t)e^{in\omega t}dt$ is  the Fourier component in the frequency space and $T$ is the period of the CPL.  In the off-resonant regime, the effective Hamiltonian of an isolated single Dirac cone is derived as $h_F(\bm{k})=(k_x\sigma_y-k_y\sigma_x)+m_{\mbox{\tiny eff}}\sigma_z$, where $m_{\mbox{\tiny eff}}=m-\eta\frac{\tilde{A}^2}{\omega}$ is the  effective mass induced by the light.

\subsection{The first-principles calculations with Floquet theory}

We performed the first-principles calculations for 3-SL and 4-SL MnBi$_2$Te$_4$  by employing the Vienna {\it ab-initio} simulation package (VASP) \cite{Kresse1996vasp, Kresse1999vasp} and the generalized gradient approximation (GGA) with the Perdew-Burke-Ernzerhof(PBE) \cite{Perdew1996PBE, Blochl1994PBE} type exchange-correlation potential is adopted, with the energy cutoff fixed to 430 eV.  By considering the transition metal Mn, GGA+U functional with U = 3 eV for Mn$-d$ orbitals for all the results in this work. The {\bf k}-point sampling grid of the Brillouin zone in the self-consistent process is a $\Gamma$-centered Monkhorst-Pack {\bf k}-point mesh of $8\times8\times1$, and a total energy tolerance $10^{-7}$eV was adopted for self-consistent convergence. To obtain the edge state of thin films MnBi$_2$Te$_4$, we employed the maximally localized Wanneir functions (MLWF)  from the first-principles calculations \cite{marzari1997maximally, souza2001maximally} to construct MLWF-based tight-binding Hamiltonians. The Mn$-d$, Te$-p$, and Bi$-p$ orbitals were initialized for MLWFs by Wannier90 \cite{Pizzi2020wannier90}. The surface density of states is obtained by the iterative Green’s function method\cite{Sancho1984Quick,Sancho1985Highly}.

Under the CPL $\bm{A}(t)$ in time with a periodicity $T$, the time-dependent Hamiltonian based on MLWFs is given by Peierls substitution,

\begin{equation}
\begin{split}
H^W(\bm{k},t)&=\sum_{\bm{R}}e^{i(\bm{k}+\frac{e}{\hbar}\bm{A}(t))\cdot(\bm{R}+\tau_n-\tau_m)}\langle m0|H|n\bm{R}\rangle,
\end{split}
\end{equation}
where $|n\bm{R}\rangle$ is the $n$-th Wannier orbital with the lattice vector $\bm{R}$, and $\tau_n$ is the center in the unit cell of the $n$-th Wannier orbital $|n0\rangle$.  The effective Hamiltonian at the high frequencies is given as, 

\begin{equation}
\begin{split}
H^{W}_{\mbox{\tiny eff}}=H^W_0+\sum_{l\ge 1}\frac{[H_{-l}^W,H_l^W]}{l \omega},
\end{split}
\end{equation}
where $H_l^W=\frac{1}{T}\int^T_0e^{il\omega t} H^W(\bm{k},t)dt$ is the Fourier transform of $H^W(\bm{k},t)$.

\section{RESULTS}
\label{results}
As shown in Fig.~\ref{fig1}(a), the stoichiometric MnBi$_2$Te$_4$ consists of SLs coupled to each other by a vdW-type interaction and features an A-type AFM in the ground state. It was revealed to be an AFM topological insulator with a gapped surface Dirac-cone state on the (111) surface\cite{zhang2019topological,li2019intrinsic}. More interestingly, MnBi$_2$Te$_4$ thin films exhibit alternating topological phases for odd- and even-SL films. In an odd-SL MnBi$_2$Te$_4$ thin films, a QAH state with zero magnetic field has been experimentally observed\cite{deng2020quantum}, and an axion insulator state, which is characterized by a zero Hall plateau, has been confirmed in even-SL thin films in recent experiments\cite{liu2020robust}. 

Then we are to reveal how the CPL engineers the topological electronic structure of MnBi$_2$Te$_4$ films. We first see the effect of the CPL on a single Dirac-cone state. The CPL induces an effective mass, written as $m_{\mbox{\tiny eff}}=m-\eta\frac{\tilde{A}^2}{\omega}$. As depicted in Fig.~\ref{fig1}(b), the RCPL can induce a process of the band gap closing and reopening with a topological transition, while the band gap monotonously increases under  LCPL irradiation. Therefore, it is straightforward to know the results of the two Dirac-cone states without coupling ($\Delta\rightarrow0$) under a CPL. When the magnetic moments of the top and bottom surfaces are along the same direction (e.g. odd-SL MnBi$_2$Te$_4$ films), the light can change the effective mass simultaneously, so the RCPL is expected to induce a sign reverse of the Chern number $|C|=1$ of the QAH state, revising the chirality of the chiral edge state. In contrast, the LCPL enhances the band gap of surface Dirac-cone states, and the QAH state remains unchanged. In addition, when the magnetic moments of the top and bottom surfaces are along the opposite directions (e.g. even-SL MnBi$_2$Te$_4$ films in the axion insulator phase with $C=0$), the CPL changes the sign of the effective mass $m_{\mbox{\tiny eff}}$ of only one (the top or the bottom, not both) surface, while the effective mass $m_{\mbox{\tiny eff}}$ on the other surface remains unchanged. Therefore, both the LCPL and the RCPL can induce the axion insulator state $C=0$ to a QAH state $|C|=1$. 

In the following, we will present the detailed results for the case of MnBi$_2$Te$_4$ films with a finite thickness and a nonzero $\Delta$, based on the effective model and the first-principles calculations.

\subsection{MnBi$_2$Te$_4$ thin films}

For MnBi$_2$Te$_4$ thin films, the two surface Dirac-cone states are coupled with a finite coupling strength ($\Delta$), which decreases as the thickness of the films increases. In the off-resonant regime, the CPL introduces an additional term $H_F$ in the effective Hamiltonian,

  \begin{equation}
  \begin{split}
  H_F=-\eta\frac{\tilde{A}^2}{\omega}I_{2}\otimes\sigma_z, 
  \end{split}
  \end{equation}
where  $\eta=\pm1$ represents the RCPL (LCPL) and $\omega$ is the frequency of the CPL.   

For the odd-SL films,  the phase diagram with the coupling strength $\Delta$ and the RCPL amplitude $\tilde{A}$ is shown in Fig.~\ref{fig2}(b). Here, we set $\omega=1$ eV and $m=50$ meV. We can see there are three phases, including two QAH phases with $C=\pm1$ and a NI phase with $C=0$. We take the 3-SL MnBi$_2$Te$_4$ film with a finite $\Delta<m$ as an example. As the light amplitude $\tilde{A}$ increases, the QAH state with the Chen number $C=+1$ in the 3-SLs MnBi$_2$Te$_4$ film can be engineered into a NI regime, and then be another QAH state with the sign reversed from $C=+1$ to $C=-1$. 

The above results from effective model analysis can be further confirmed by our first-principles calculations. Based on first-principles calculations, the band gap evolution with varying $\tilde{A}$ of the RCPL and LCPL are calculated and shown in Figs.~\ref{fig2}(c) and 2(d), respectively. For the RCPL, the band gap clearly undergoes the process of closing and reopening with increasing the light amplitude $\tilde{A}$, which indicates that the topological transition occurs at the band gap closing points. Differently, the LCPL only enhances the band gap, as shown in Fig.~\ref{fig2}(d), which provides an interesting way to raise the working temperature of the QAH effect. To further demonstrate the topological transitions as $\tilde{A}$ increases for RCPL , as shown in Fig.~\ref{fig2}(c), we calculate the surface states in Fig.~\ref{fig2}(e, f, g) for $\tilde{A}=0.06$ \AA$^{-1}$, $0.11$ \AA$^{-1}$ and $0.13$ \AA$^{-1}$, respectively. The chiral edge state appears at $\tilde{A}=0.06$ \AA$^{-1}$, disappears at $\tilde{A}=0.11$ \AA$^{-1}$, and changes its chirality at $\tilde{A}=0.13$ \AA$^{-1}$. Therefore, our first-principles calculations are consistent with the above effective model analysis.  We also can convert the amplitude of the vector potential to electric field strength as $E=-\partial A/\partial t\sim-A\omega$. For the frequency of $\omega=1 eV$ used in this paper, a typical value $\tilde{A}=0.11$ \AA$^{-1}$ corresponds to the electric field strength as $0.11 V/$\AA, which should be within the experimental feasibility\cite{Wang2013Observation,hubener2017creating}.

 As mentioned above, even-SL MnBi$_2$Te$_4$ films are in an axion insulator phase with a zero Hall plateau ($C=0$). The Zeeman coupling $m_1=-m_2=m$ describes the opposite magnetic moment in the top and bottom surfaces in the effective Hamiltonian (Eq.~\ref{eq1}) for the even-SL MnBi$_2$Te$_4$ thin films. Figure~\ref{fig3}(b) shows the phase diagram with the coupling strength $\Delta$ and the RCPL/LCPL amplitude $\tilde{A}$. We can see that there is always a topological transition with increasing the light amplitude $\tilde{A}$ for both the RCPL and LCPL regardless of the coupling strength $\Delta$.  We take 4-SL MnBi$_2$Te$_4$ films as an example in our first-principles calculations, the band gap evolution on the light amplitude $\tilde{A}$ is calculated and shown in Fig.~\ref{fig3}(c). We can see that the band gap first decreases until it closes and then reopens again, along with the change of the axion insulator phase into a QAH phase with $C=\pm1$. We also calculate the edge state of 4-SL MnBi$_2$Te$_4$ film under the RCPL/LCPL by first-principles calculations. As expected, there is no chiral edge state at $\tilde{A}=0.06$ \AA$^{-1}$ in the axion insulator phase, as seen in Fig.~\ref{fig3}(d). When the light amplitude increases to, e.g., $\tilde{A}=0.012$ \AA$^{-1}$, the chiral edge state with the opposite chiralities for the RCPL and LCPL arises, as seen in Fig.~\ref{fig3}(e, f). Therefore, both the effective model analysis and first-principles calculations reveal that the light in the off-resonant regime can drive the axion insulator into a QAH state, and the different handedness of the CPL can induce the QAH state with different chiralities of the chiral edge state.

\begin{figure}[t]
  \centering
  \includegraphics[width=3.4in]{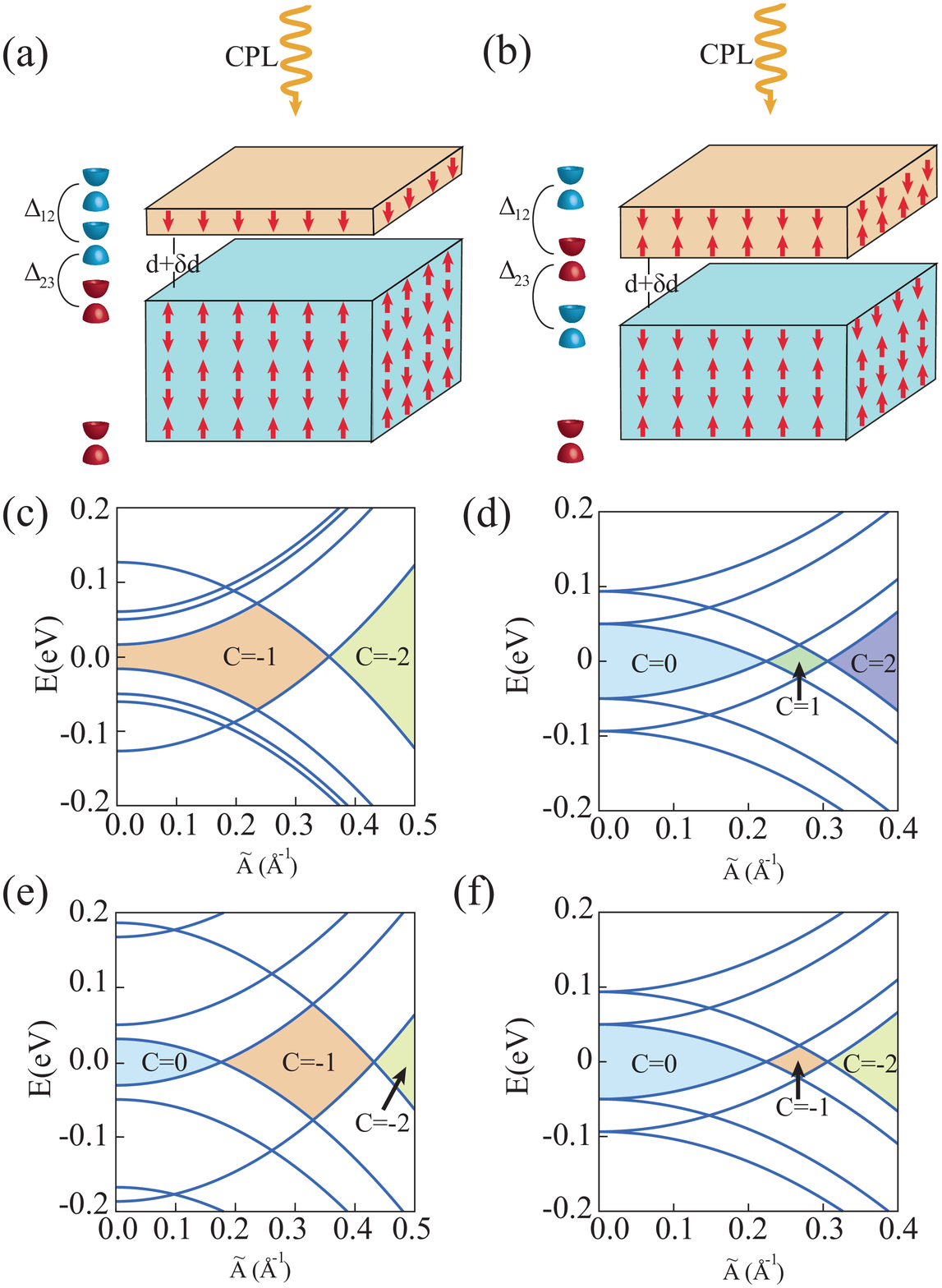}\
  \caption{ The schematic diagram of the CPL shines on the thick AFM MnBi$_2$Te$_4$ films with tunable  (a) FM surface layer and (b) AFM surface layer. The surface state is described by four coupled massive Dirac cones, the label of the Dirac cones increases from top to the bottom. The  surface vdW gap is $d+\delta d$.  The energy at $\bm{k}=0$ vs $\tilde{A}$ of a FM surface layer under the irradiation of LCPL with (c) $\Delta_{23}=25$ meV and (e) $\Delta_{23}=150$ meV. The energy at $\bm{k}=0$ vs $\tilde{A}$ of a AFM surface layer under the irradiation of (d) RCPL  and (f) LCPL with $\Delta_{23}=25$ meV The parameters are $\Delta_{12}=75$ meV,  $m=50$ meV, $\omega=1$ eV.   }\label{fig4}
\end{figure}

\subsection{MnBi$_2$Te$_4$ thick films with a surface vdW gap expansion}

 In the MnBi$_2$Te$_4$ thick films, the coupling strength $\Delta\rightarrow0$ between the top and bottom Dirac-cone surface states. The phase transition under the irradiation of CPL can be directly understood by the sum of two isolated Dirac cones.  However, because MnBi$_2$Te$_4$ SLs are coupled through the weak vdW interaction, the topmost vdW gap on the surface is expected to slightly expand ($\approx0.2$ \AA) due to the cleave processing, the temperature or the impurities on the surface\cite{wang2022three}. This slight expansion is predicted to drastically reduce, or even close, the band gap of the surface Dirac-cone state of MnBi$_2$Te$_4$, which provides a new point of view to understand the observed (nearly) gapless surface states in many angle resolved photoemission spectroscopy (ARPES) experiments\cite{gong2019experimental,he2020mnbi,chen2019topological,hao2019gapless}. Here, we consider two cases of the expansion of the surface vdW gap, as depicted in Fig.~\ref{fig4}(a, b). The surface vdW gap between a top ferromagnetic (FM) SL or AFM bi-SL and the below bulk is $d+\delta d$ where $d$ is the bulk vdW gap, and $\delta d$ is the expansion of the surface vdW gap. The low-energy effective model to describe the expansion of the surface vdW gap of MnBi$_2$Te$_4$ films can be written as a four-Dirac-fermion model,

\begin{equation}
\label{4D}
\begin{split}
H_{4D}(\bm{k})=\begin{pmatrix}
h_1(\bm{k})&\Delta_{12}I_{2}&0&0\\
\Delta_{12}I_{2}&h_2(\bm{k})&\Delta_{23}I_{2}&0\\
0&\Delta_{23}I_{2}&h_3(\bm{k})&\Delta_{34}I_{2}\\
0&0&\Delta_{34}I_{2}&h_4(\bm{k})
\end{pmatrix}
\end{split}
\end{equation}
where $h_i(\bm{k})$ is the $i$-th Dirac cone with $i=1,2,3,4$. The first and the second Dirac cones are on the top and bottom surfaces of the top surface layer, and the third and fourth Dirac cones are on the top and bottom surfaces of the below part of the MnBi$_2$Te$_4$ thick film (e.g. 6-SL film), as schematically shown in Fig.~\ref{fig4}(a,b). $\Delta_{ij}$ is the coupling strength of the $i$-th and $j$-th Dirac cones, $s_i=+1$ ($-1$) for $i$=1, 3 (2, 4). If we consider the FM surface SL as shown in Fig.~\ref{fig4}(a), we set $m_1=m_2=-m_3=-m_4=m$ for the model in Eq.~\ref{4D}, and for an AFM surface bi-SL, we have $m_1=-m_2=m_3=-m_4=m$. The surface layer is thin enough to have $\Delta_{12}>h$. In addition, the coupling strength $D_{34}\rightarrow 0$ in a thick film (e.g. 6-SL film). 

When we apply a CPL along the [001] direction to the above 6-SL MnBi$_2$Te$_4$ thick film, the effective Hamiltonian obtains an additional term written as $H_{4D}^F=-\eta\frac{\tilde{A}^2}{\omega}I_{4}\otimes\sigma_z$,  where $I_{4}$ is the $4\times 4$ identity matrix. In Fig.~\ref{fig4}(c, e), we show the calculated band gap evolution at $\Gamma$ with varying $\tilde{A}$ for the FM surface layer under a LCPL. It is clear that the Chern number $C=-1$ in the decoupling limit ($\Delta_{23}=0$) for the 6-SL film without the light. As an example, we set a finite $\Delta_{23}=25$ meV in Fig.~\ref{fig4}(c). We can see that the LCPL can drive $C=-1$ to $-2$ as the energy gap undergoes a process of closing and reopening. When the coupling strength $\Delta_{23}$ increases but still does not reach the level of the coupling strength in the bulk, for example, $\Delta_{23}=150$ meV in  Fig.~\ref{fig4}(e). We can still obtain a high Chern number $C=-2$ when the CPL amplitude is large enough.

When the surface layer is an AFM bi-SL, a QAH state with a high Chern number $|C|=2$ also arises. We depict the calculated band gap at $\Gamma$ versus $\tilde{A}$ under the RCPL (LCPL) in Fig.~\ref{fig4}(d, f), respectively. We can see that the band gap goes through the process of closing and reopening twice. With a large $\tilde{A}$, a high Chern number $|C|=2$ can arise for both the RCPL and LCPL, but the chirality of the edge states is opposite. At the decoupling limit of $\Delta_{23}\rightarrow 0$, the system can be seen as two axion insulators without coupling. As discussed above, the CPL can drive an axion insulator into a Chern insulator with $|C|=1$ and the sign of the Chern number corresponds to the hand of the CPL, so if we stack two axion insulators without coupling, $|C|=2$ is reasonable to be understood. As $\Delta_{23}$ slowly increases, we can still  keep $|C|=2$ but requiring a large amplitude of the light. Therefore, different from previous works\cite{zhao2020tuning,Zhao2022Zero,ge2020high}, we provide an alternative way to obtain a high-Chern-number QAH state in MnBi$_2$Te$_4$ films through the light.

\section{Conclusion}
\label{conclusion}
To summarize, utilizing the effective model and first-principles calculations in combination with Floquet theory, our work shows that the CPL can induce rich topological phase transitions in AFM topological insulator MnBi$_2$Te$_4$ films. For odd-SL MnBi$_2$Te$_4$ thin films, the RCPL can reverse the Chern number from $C=1$ to $C=-1$ for the QAH state, with simultaneously reversing the chirality of the chiral edge state. The LCPL can enhance the band gap of the QAH state of the odd-SL MnBi$_2$Te$_4$ thin films, which provides a convenient way to improve its working temperature. For even-SL MnBi$_2$Te$_4$ thin films, the CPL can drive the axion insulator into a QAH state, and interestingly, the CPL with the different handedness induces chiral edge states with opposite chiralities. Finally, we further discuss the case of the CPL on the MnBi$_2$Te$_4$ thick films with a slight expansion of the surface vdW gap. The QAH states with a high Chern number $C=\pm2$ can be driven by the CPL when the amplitude of the light is large enough. This is an alternative way to achieve a high Chern number QAH effect, different from previous works, such as, an external magnetic field and the physically stacking QAH layers. Our results reveal the CPL-driven topological phase transitions in MnBi$_2$Te$_4$ films, and may have potential applications for the design of optoelectronic devices in the future.

\section*{ACKNOWLEDGEMENTS}
This work is supported by National Key Projects for Research and Development of China (Grant No.2021YFA1400400), the Fundamental Research Funds for the Central Universities (Grant No. 020414380185), Natural Science Foundation of Jiangsu Province (No. BK20200007), the Natural Science Foundation of China (No. 12074181, No. 11834006, and No. 12104217) and the Fok Ying-Tong Education Foundation of China (Grant No. 161006).

T. Z and H. W contributed equally to this work.
\bibliography{ref}

\begin{thebibliography}{73}
\expandafter\ifx\csname natexlab\endcsname\relax\def\natexlab#1{#1}\fi
\expandafter\ifx\csname bibnamefont\endcsname\relax
  \def\bibnamefont#1{#1}\fi
\expandafter\ifx\csname bibfnamefont\endcsname\relax
  \def\bibfnamefont#1{#1}\fi
\expandafter\ifx\csname citenamefont\endcsname\relax
  \def\citenamefont#1{#1}\fi
\expandafter\ifx\csname url\endcsname\relax
  \def\url#1{\texttt{#1}}\fi
\expandafter\ifx\csname urlprefix\endcsname\relax\def\urlprefix{URL }\fi
\providecommand{\bibinfo}[2]{#2}
\providecommand{\eprint}[2][]{\url{#2}}

\bibitem[{\citenamefont{Qi and Zhang}(2011)}]{Qi2011Topological}
\bibinfo{author}{\bibfnamefont{X.-L.} \bibnamefont{Qi}} \bibnamefont{and}
  \bibinfo{author}{\bibfnamefont{S.-C.} \bibnamefont{Zhang}},
  \bibinfo{journal}{Rev. Mod. Phys.} \textbf{\bibinfo{volume}{83}},
  \bibinfo{pages}{1057} (\bibinfo{year}{2011}).

\bibitem[{\citenamefont{Hasan and Kane}(2010)}]{Hasan2010Colloquium}
\bibinfo{author}{\bibfnamefont{M.~Z.} \bibnamefont{Hasan}} \bibnamefont{and}
  \bibinfo{author}{\bibfnamefont{C.~L.} \bibnamefont{Kane}},
  \bibinfo{journal}{Rev. Mod. Phys.} \textbf{\bibinfo{volume}{82}},
  \bibinfo{pages}{3045} (\bibinfo{year}{2010}).

\bibitem[{\citenamefont{Haldane}(1988)}]{Haldane1988}
\bibinfo{author}{\bibfnamefont{F.~D.~M.} \bibnamefont{Haldane}},
  \bibinfo{journal}{Phys. Rev. Lett.} \textbf{\bibinfo{volume}{61}},
  \bibinfo{pages}{2015} (\bibinfo{year}{1988}).

\bibitem[{\citenamefont{Qi et~al.}(2008)\citenamefont{Qi, Hughes, and
  Zhang}}]{qi2008}
\bibinfo{author}{\bibfnamefont{X.-L.} \bibnamefont{Qi}},
  \bibinfo{author}{\bibfnamefont{T.~L.} \bibnamefont{Hughes}},
  \bibnamefont{and} \bibinfo{author}{\bibfnamefont{S.-C.} \bibnamefont{Zhang}},
  \bibinfo{journal}{Phys. Rev. B} \textbf{\bibinfo{volume}{78}},
  \bibinfo{pages}{195424} (\bibinfo{year}{2008}).

\bibitem[{\citenamefont{Li et~al.}(2010)\citenamefont{Li, Wang, Qi, and
  Zhang}}]{li2010}
\bibinfo{author}{\bibfnamefont{R.}~\bibnamefont{Li}},
  \bibinfo{author}{\bibfnamefont{J.}~\bibnamefont{Wang}},
  \bibinfo{author}{\bibfnamefont{X.~L.} \bibnamefont{Qi}}, \bibnamefont{and}
  \bibinfo{author}{\bibfnamefont{S.~C.} \bibnamefont{Zhang}},
  \bibinfo{journal}{Nat. Phys.} \textbf{\bibinfo{volume}{6}},
  \bibinfo{pages}{284} (\bibinfo{year}{2010}).

\bibitem[{\citenamefont{Yu et~al.}(2010)\citenamefont{Yu, Zhang, Zhang, Zhang,
  Dai, and Fang}}]{yu2010}
\bibinfo{author}{\bibfnamefont{R.}~\bibnamefont{Yu}},
  \bibinfo{author}{\bibfnamefont{W.}~\bibnamefont{Zhang}},
  \bibinfo{author}{\bibfnamefont{H.-J.} \bibnamefont{Zhang}},
  \bibinfo{author}{\bibfnamefont{S.-C.} \bibnamefont{Zhang}},
  \bibinfo{author}{\bibfnamefont{X.}~\bibnamefont{Dai}}, \bibnamefont{and}
  \bibinfo{author}{\bibfnamefont{Z.}~\bibnamefont{Fang}},
  \bibinfo{journal}{Science} \textbf{\bibinfo{volume}{329}},
  \bibinfo{pages}{61} (\bibinfo{year}{2010}).

\bibitem[{\citenamefont{Chang et~al.}(2013)\citenamefont{Chang, Zhang, Feng,
  Shen, Zhang, Guo, Li, Ou, Wei, Wang et~al.}}]{chang2013experimental}
\bibinfo{author}{\bibfnamefont{C.-Z.} \bibnamefont{Chang}},
  \bibinfo{author}{\bibfnamefont{J.}~\bibnamefont{Zhang}},
  \bibinfo{author}{\bibfnamefont{X.}~\bibnamefont{Feng}},
  \bibinfo{author}{\bibfnamefont{J.}~\bibnamefont{Shen}},
  \bibinfo{author}{\bibfnamefont{Z.}~\bibnamefont{Zhang}},
  \bibinfo{author}{\bibfnamefont{M.}~\bibnamefont{Guo}},
  \bibinfo{author}{\bibfnamefont{K.}~\bibnamefont{Li}},
  \bibinfo{author}{\bibfnamefont{Y.}~\bibnamefont{Ou}},
  \bibinfo{author}{\bibfnamefont{P.}~\bibnamefont{Wei}},
  \bibinfo{author}{\bibfnamefont{L.-L.} \bibnamefont{Wang}},
  \bibnamefont{et~al.}, \bibinfo{journal}{Science}
  \textbf{\bibinfo{volume}{340}}, \bibinfo{pages}{167} (\bibinfo{year}{2013}).

\bibitem[{\citenamefont{Gong et~al.}(2019)\citenamefont{Gong, Guo, Li, Zhu,
  Liao, Liu, Zhang, Gu, Tang, Feng et~al.}}]{gong2019experimental}
\bibinfo{author}{\bibfnamefont{Y.}~\bibnamefont{Gong}},
  \bibinfo{author}{\bibfnamefont{J.}~\bibnamefont{Guo}},
  \bibinfo{author}{\bibfnamefont{J.}~\bibnamefont{Li}},
  \bibinfo{author}{\bibfnamefont{K.}~\bibnamefont{Zhu}},
  \bibinfo{author}{\bibfnamefont{M.}~\bibnamefont{Liao}},
  \bibinfo{author}{\bibfnamefont{X.}~\bibnamefont{Liu}},
  \bibinfo{author}{\bibfnamefont{Q.}~\bibnamefont{Zhang}},
  \bibinfo{author}{\bibfnamefont{L.}~\bibnamefont{Gu}},
  \bibinfo{author}{\bibfnamefont{L.}~\bibnamefont{Tang}},
  \bibinfo{author}{\bibfnamefont{X.}~\bibnamefont{Feng}}, \bibnamefont{et~al.},
  \bibinfo{journal}{Chin. Phys. Lett.} \textbf{\bibinfo{volume}{36}},
  \bibinfo{pages}{076801} (\bibinfo{year}{2019}).

\bibitem[{\citenamefont{Zhang et~al.}(2019)\citenamefont{Zhang, Shi, Zhu, Xing,
  Zhang, and Wang}}]{zhang2019topological}
\bibinfo{author}{\bibfnamefont{D.}~\bibnamefont{Zhang}},
  \bibinfo{author}{\bibfnamefont{M.}~\bibnamefont{Shi}},
  \bibinfo{author}{\bibfnamefont{T.}~\bibnamefont{Zhu}},
  \bibinfo{author}{\bibfnamefont{D.}~\bibnamefont{Xing}},
  \bibinfo{author}{\bibfnamefont{H.}~\bibnamefont{Zhang}}, \bibnamefont{and}
  \bibinfo{author}{\bibfnamefont{J.}~\bibnamefont{Wang}},
  \bibinfo{journal}{Phys. Rev. Lett.} \textbf{\bibinfo{volume}{122}},
  \bibinfo{pages}{206401} (\bibinfo{year}{2019}).

\bibitem[{\citenamefont{Li et~al.}(2019{\natexlab{a}})\citenamefont{Li, Li, Du,
  Wang, Gu, Zhang, He, Duan, and Xu}}]{li2019intrinsic}
\bibinfo{author}{\bibfnamefont{J.}~\bibnamefont{Li}},
  \bibinfo{author}{\bibfnamefont{Y.}~\bibnamefont{Li}},
  \bibinfo{author}{\bibfnamefont{S.}~\bibnamefont{Du}},
  \bibinfo{author}{\bibfnamefont{Z.}~\bibnamefont{Wang}},
  \bibinfo{author}{\bibfnamefont{B.-L.} \bibnamefont{Gu}},
  \bibinfo{author}{\bibfnamefont{S.-C.} \bibnamefont{Zhang}},
  \bibinfo{author}{\bibfnamefont{K.}~\bibnamefont{He}},
  \bibinfo{author}{\bibfnamefont{W.}~\bibnamefont{Duan}}, \bibnamefont{and}
  \bibinfo{author}{\bibfnamefont{Y.}~\bibnamefont{Xu}}, \bibinfo{journal}{Sci.
  Adv.} \textbf{\bibinfo{volume}{5}}, \bibinfo{pages}{eaaw5685}
  (\bibinfo{year}{2019}{\natexlab{a}}).

\bibitem[{\citenamefont{Otrokov et~al.}(2019)\citenamefont{Otrokov,
  Klimovskikh, Bentmann, Estyunin, Zeugner, Aliev, Ga{\ss}, Wolter, Koroleva,
  Shikin et~al.}}]{otrokov2019prediction}
\bibinfo{author}{\bibfnamefont{M.~M.} \bibnamefont{Otrokov}},
  \bibinfo{author}{\bibfnamefont{I.~I.} \bibnamefont{Klimovskikh}},
  \bibinfo{author}{\bibfnamefont{H.}~\bibnamefont{Bentmann}},
  \bibinfo{author}{\bibfnamefont{D.}~\bibnamefont{Estyunin}},
  \bibinfo{author}{\bibfnamefont{A.}~\bibnamefont{Zeugner}},
  \bibinfo{author}{\bibfnamefont{Z.~S.} \bibnamefont{Aliev}},
  \bibinfo{author}{\bibfnamefont{S.}~\bibnamefont{Ga{\ss}}},
  \bibinfo{author}{\bibfnamefont{A.}~\bibnamefont{Wolter}},
  \bibinfo{author}{\bibfnamefont{A.}~\bibnamefont{Koroleva}},
  \bibinfo{author}{\bibfnamefont{A.~M.} \bibnamefont{Shikin}},
  \bibnamefont{et~al.}, \bibinfo{journal}{Nature}
  \textbf{\bibinfo{volume}{576}}, \bibinfo{pages}{416} (\bibinfo{year}{2019}).

\bibitem[{\citenamefont{Chen et~al.}(2019{\natexlab{a}})\citenamefont{Chen, Xu,
  Li, Li, Wang, Zhang, Li, Wu, Liang, Chen et~al.}}]{chen2019topological}
\bibinfo{author}{\bibfnamefont{Y.~J.} \bibnamefont{Chen}},
  \bibinfo{author}{\bibfnamefont{L.~X.} \bibnamefont{Xu}},
  \bibinfo{author}{\bibfnamefont{J.~H.} \bibnamefont{Li}},
  \bibinfo{author}{\bibfnamefont{Y.~W.} \bibnamefont{Li}},
  \bibinfo{author}{\bibfnamefont{H.~Y.} \bibnamefont{Wang}},
  \bibinfo{author}{\bibfnamefont{C.~F.} \bibnamefont{Zhang}},
  \bibinfo{author}{\bibfnamefont{H.}~\bibnamefont{Li}},
  \bibinfo{author}{\bibfnamefont{Y.}~\bibnamefont{Wu}},
  \bibinfo{author}{\bibfnamefont{A.~J.} \bibnamefont{Liang}},
  \bibinfo{author}{\bibfnamefont{C.}~\bibnamefont{Chen}}, \bibnamefont{et~al.},
  \bibinfo{journal}{Phys. Rev. X} \textbf{\bibinfo{volume}{9}},
  \bibinfo{pages}{041040} (\bibinfo{year}{2019}{\natexlab{a}}).

\bibitem[{\citenamefont{Hao et~al.}(2019)\citenamefont{Hao, Liu, Feng, Ma,
  Schwier, Arita, Kumar, Hu, Lu, Zeng et~al.}}]{hao2019gapless}
\bibinfo{author}{\bibfnamefont{Y.-J.} \bibnamefont{Hao}},
  \bibinfo{author}{\bibfnamefont{P.}~\bibnamefont{Liu}},
  \bibinfo{author}{\bibfnamefont{Y.}~\bibnamefont{Feng}},
  \bibinfo{author}{\bibfnamefont{X.-M.} \bibnamefont{Ma}},
  \bibinfo{author}{\bibfnamefont{E.~F.} \bibnamefont{Schwier}},
  \bibinfo{author}{\bibfnamefont{M.}~\bibnamefont{Arita}},
  \bibinfo{author}{\bibfnamefont{S.}~\bibnamefont{Kumar}},
  \bibinfo{author}{\bibfnamefont{C.}~\bibnamefont{Hu}},
  \bibinfo{author}{\bibfnamefont{R.}~\bibnamefont{Lu}},
  \bibinfo{author}{\bibfnamefont{M.}~\bibnamefont{Zeng}}, \bibnamefont{et~al.},
  \bibinfo{journal}{Phys. Rev. X} \textbf{\bibinfo{volume}{9}},
  \bibinfo{pages}{041038} (\bibinfo{year}{2019}).

\bibitem[{\citenamefont{Li et~al.}(2019{\natexlab{b}})\citenamefont{Li, Gao,
  Duan, Xu, Zhu, Tian, Gao, Fan, Rao, Huang et~al.}}]{li2019gapless}
\bibinfo{author}{\bibfnamefont{H.}~\bibnamefont{Li}},
  \bibinfo{author}{\bibfnamefont{S.-Y.} \bibnamefont{Gao}},
  \bibinfo{author}{\bibfnamefont{S.-F.} \bibnamefont{Duan}},
  \bibinfo{author}{\bibfnamefont{Y.-F.} \bibnamefont{Xu}},
  \bibinfo{author}{\bibfnamefont{K.-J.} \bibnamefont{Zhu}},
  \bibinfo{author}{\bibfnamefont{S.-J.} \bibnamefont{Tian}},
  \bibinfo{author}{\bibfnamefont{J.-C.} \bibnamefont{Gao}},
  \bibinfo{author}{\bibfnamefont{W.-H.} \bibnamefont{Fan}},
  \bibinfo{author}{\bibfnamefont{Z.-C.} \bibnamefont{Rao}},
  \bibinfo{author}{\bibfnamefont{J.-R.} \bibnamefont{Huang}},
  \bibnamefont{et~al.}, \bibinfo{journal}{Phys. Rev. X}
  \textbf{\bibinfo{volume}{9}}, \bibinfo{pages}{041039}
  (\bibinfo{year}{2019}{\natexlab{b}}).

\bibitem[{\citenamefont{E.~D. L.~Rienks et~al.}(2019)\citenamefont{E.~D.
  L.~Rienks, S{\'a}nchez-Barriga et~al.}}]{rienks2019large}
\bibinfo{author}{\bibfnamefont{J.~S.} \bibnamefont{E.~D. L.~Rienks},
  \bibfnamefont{S.~Wimmer}}, \bibinfo{author}{\bibfnamefont{P.~S. M. J. A. H.
  V. H. M. A. G.~K.} \bibnamefont{S{\'a}nchez-Barriga},
  \bibfnamefont{O.~Caha}}, \bibnamefont{et~al.}, \bibinfo{journal}{Nature}
  \textbf{\bibinfo{volume}{576}}, \bibinfo{pages}{423} (\bibinfo{year}{2019}).

\bibitem[{\citenamefont{Chen et~al.}(2019{\natexlab{b}})\citenamefont{Chen,
  Fei, Zhang, Zhang, Liu, Zhang, Wang, Wei, Zhang, Zuo
  et~al.}}]{chen2019intrinsic}
\bibinfo{author}{\bibfnamefont{B.}~\bibnamefont{Chen}},
  \bibinfo{author}{\bibfnamefont{F.}~\bibnamefont{Fei}},
  \bibinfo{author}{\bibfnamefont{D.}~\bibnamefont{Zhang}},
  \bibinfo{author}{\bibfnamefont{B.}~\bibnamefont{Zhang}},
  \bibinfo{author}{\bibfnamefont{W.}~\bibnamefont{Liu}},
  \bibinfo{author}{\bibfnamefont{S.}~\bibnamefont{Zhang}},
  \bibinfo{author}{\bibfnamefont{P.}~\bibnamefont{Wang}},
  \bibinfo{author}{\bibfnamefont{B.}~\bibnamefont{Wei}},
  \bibinfo{author}{\bibfnamefont{Y.}~\bibnamefont{Zhang}},
  \bibinfo{author}{\bibfnamefont{Z.}~\bibnamefont{Zuo}}, \bibnamefont{et~al.},
  \bibinfo{journal}{Nat. Commun.} \textbf{\bibinfo{volume}{10}},
  \bibinfo{pages}{4468} (\bibinfo{year}{2019}{\natexlab{b}}).

\bibitem[{\citenamefont{Vidal et~al.}(2019)\citenamefont{Vidal, Zeugner, Facio,
  Ray, Haghighi, Wolter, Bohorquez, Caglieris, Moser, Figgemeier
  et~al.}}]{vidal2019topological}
\bibinfo{author}{\bibfnamefont{R.~C.} \bibnamefont{Vidal}},
  \bibinfo{author}{\bibfnamefont{A.}~\bibnamefont{Zeugner}},
  \bibinfo{author}{\bibfnamefont{J.~I.} \bibnamefont{Facio}},
  \bibinfo{author}{\bibfnamefont{R.}~\bibnamefont{Ray}},
  \bibinfo{author}{\bibfnamefont{M.~H.} \bibnamefont{Haghighi}},
  \bibinfo{author}{\bibfnamefont{A.~U.} \bibnamefont{Wolter}},
  \bibinfo{author}{\bibfnamefont{L.~T.~C.} \bibnamefont{Bohorquez}},
  \bibinfo{author}{\bibfnamefont{F.}~\bibnamefont{Caglieris}},
  \bibinfo{author}{\bibfnamefont{S.}~\bibnamefont{Moser}},
  \bibinfo{author}{\bibfnamefont{T.}~\bibnamefont{Figgemeier}},
  \bibnamefont{et~al.}, \bibinfo{journal}{Phys. Rev. X}
  \textbf{\bibinfo{volume}{9}}, \bibinfo{pages}{041065} (\bibinfo{year}{2019}).

\bibitem[{\citenamefont{Klimovskikh et~al.}(2020)\citenamefont{Klimovskikh,
  Otrokov, Estyunin, Eremeev, Filnov, Koroleva, Shevchenko, Voroshnin, Rybkin,
  Rusinov et~al.}}]{klimovskikh2020tunable}
\bibinfo{author}{\bibfnamefont{I.~I.} \bibnamefont{Klimovskikh}},
  \bibinfo{author}{\bibfnamefont{M.~M.} \bibnamefont{Otrokov}},
  \bibinfo{author}{\bibfnamefont{D.}~\bibnamefont{Estyunin}},
  \bibinfo{author}{\bibfnamefont{S.~V.} \bibnamefont{Eremeev}},
  \bibinfo{author}{\bibfnamefont{S.~O.} \bibnamefont{Filnov}},
  \bibinfo{author}{\bibfnamefont{A.}~\bibnamefont{Koroleva}},
  \bibinfo{author}{\bibfnamefont{E.}~\bibnamefont{Shevchenko}},
  \bibinfo{author}{\bibfnamefont{V.}~\bibnamefont{Voroshnin}},
  \bibinfo{author}{\bibfnamefont{A.~G.} \bibnamefont{Rybkin}},
  \bibinfo{author}{\bibfnamefont{I.~P.} \bibnamefont{Rusinov}},
  \bibnamefont{et~al.}, \bibinfo{journal}{npj Quantum Mater.}
  \textbf{\bibinfo{volume}{5}}, \bibinfo{pages}{54} (\bibinfo{year}{2020}).

\bibitem[{\citenamefont{Zhu et~al.}(2021{\natexlab{a}})\citenamefont{Zhu, Wang,
  Zhang, and Xing}}]{zhu2021tunable}
\bibinfo{author}{\bibfnamefont{T.}~\bibnamefont{Zhu}},
  \bibinfo{author}{\bibfnamefont{H.}~\bibnamefont{Wang}},
  \bibinfo{author}{\bibfnamefont{H.}~\bibnamefont{Zhang}}, \bibnamefont{and}
  \bibinfo{author}{\bibfnamefont{D.}~\bibnamefont{Xing}}, \bibinfo{journal}{npj
  Comput. Mater.} \textbf{\bibinfo{volume}{7}}, \bibinfo{pages}{121}
  (\bibinfo{year}{2021}{\natexlab{a}}).

\bibitem[{\citenamefont{Wang et~al.}(2020)\citenamefont{Wang, Wang, Yang, Shi,
  Ruan, Xing, Wang, and Zhang}}]{wang2020heterostructures}
\bibinfo{author}{\bibfnamefont{H.}~\bibnamefont{Wang}},
  \bibinfo{author}{\bibfnamefont{D.}~\bibnamefont{Wang}},
  \bibinfo{author}{\bibfnamefont{Z.}~\bibnamefont{Yang}},
  \bibinfo{author}{\bibfnamefont{M.}~\bibnamefont{Shi}},
  \bibinfo{author}{\bibfnamefont{J.}~\bibnamefont{Ruan}},
  \bibinfo{author}{\bibfnamefont{D.}~\bibnamefont{Xing}},
  \bibinfo{author}{\bibfnamefont{J.}~\bibnamefont{Wang}}, \bibnamefont{and}
  \bibinfo{author}{\bibfnamefont{H.}~\bibnamefont{Zhang}},
  \bibinfo{journal}{Phys. Rev. B} \textbf{\bibinfo{volume}{101}},
  \bibinfo{pages}{081109} (\bibinfo{year}{2020}).

\bibitem[{\citenamefont{Zhang et~al.}(2020)\citenamefont{Zhang, Wang, Shi, Zhu,
  Zhang, and Wang}}]{zhang2020MBT225}
\bibinfo{author}{\bibfnamefont{J.}~\bibnamefont{Zhang}},
  \bibinfo{author}{\bibfnamefont{D.}~\bibnamefont{Wang}},
  \bibinfo{author}{\bibfnamefont{M.}~\bibnamefont{Shi}},
  \bibinfo{author}{\bibfnamefont{T.}~\bibnamefont{Zhu}},
  \bibinfo{author}{\bibfnamefont{H.}~\bibnamefont{Zhang}}, \bibnamefont{and}
  \bibinfo{author}{\bibfnamefont{J.}~\bibnamefont{Wang}},
  \bibinfo{journal}{Chin. Phys. Lett.} \textbf{\bibinfo{volume}{37}},
  \bibinfo{pages}{077304} (\bibinfo{year}{2020}).

\bibitem[{\citenamefont{Yang and Zhang}(2022)}]{Yang2022Evolution}
\bibinfo{author}{\bibfnamefont{Z.}~\bibnamefont{Yang}} \bibnamefont{and}
  \bibinfo{author}{\bibfnamefont{H.}~\bibnamefont{Zhang}},
  \bibinfo{journal}{New J. Phys.} \textbf{\bibinfo{volume}{24}},
  \bibinfo{pages}{073034} (\bibinfo{year}{2022}).

\bibitem[{\citenamefont{Zhan et~al.}(2021)\citenamefont{Zhan, Shi, Yang, and
  Zhang}}]{Zhan2021A}
\bibinfo{author}{\bibfnamefont{G.}~\bibnamefont{Zhan}},
  \bibinfo{author}{\bibfnamefont{M.}~\bibnamefont{Shi}},
  \bibinfo{author}{\bibfnamefont{Z.}~\bibnamefont{Yang}}, \bibnamefont{and}
  \bibinfo{author}{\bibfnamefont{H.}~\bibnamefont{Zhang}},
  \bibinfo{journal}{Chin. Phys. Lett.} \textbf{\bibinfo{volume}{38}},
  \bibinfo{pages}{077105} (\bibinfo{year}{2021}).

\bibitem[{\citenamefont{Deng et~al.}(2020)\citenamefont{Deng, Yu, Shi, Guo, Xu,
  Wang, Chen, and Zhang}}]{deng2020quantum}
\bibinfo{author}{\bibfnamefont{Y.}~\bibnamefont{Deng}},
  \bibinfo{author}{\bibfnamefont{Y.}~\bibnamefont{Yu}},
  \bibinfo{author}{\bibfnamefont{M.~Z.} \bibnamefont{Shi}},
  \bibinfo{author}{\bibfnamefont{Z.}~\bibnamefont{Guo}},
  \bibinfo{author}{\bibfnamefont{Z.}~\bibnamefont{Xu}},
  \bibinfo{author}{\bibfnamefont{J.}~\bibnamefont{Wang}},
  \bibinfo{author}{\bibfnamefont{X.~H.} \bibnamefont{Chen}}, \bibnamefont{and}
  \bibinfo{author}{\bibfnamefont{Y.}~\bibnamefont{Zhang}},
  \bibinfo{journal}{Science} \textbf{\bibinfo{volume}{367}},
  \bibinfo{pages}{895} (\bibinfo{year}{2020}).

\bibitem[{\citenamefont{Liu et~al.}(2020)\citenamefont{Liu, Wang, Li, Wu, Li,
  Li, He, Xu, Zhang, and Wang}}]{liu2020robust}
\bibinfo{author}{\bibfnamefont{C.}~\bibnamefont{Liu}},
  \bibinfo{author}{\bibfnamefont{Y.}~\bibnamefont{Wang}},
  \bibinfo{author}{\bibfnamefont{H.}~\bibnamefont{Li}},
  \bibinfo{author}{\bibfnamefont{Y.}~\bibnamefont{Wu}},
  \bibinfo{author}{\bibfnamefont{Y.}~\bibnamefont{Li}},
  \bibinfo{author}{\bibfnamefont{J.}~\bibnamefont{Li}},
  \bibinfo{author}{\bibfnamefont{K.}~\bibnamefont{He}},
  \bibinfo{author}{\bibfnamefont{Y.}~\bibnamefont{Xu}},
  \bibinfo{author}{\bibfnamefont{J.}~\bibnamefont{Zhang}}, \bibnamefont{and}
  \bibinfo{author}{\bibfnamefont{Y.}~\bibnamefont{Wang}},
  \bibinfo{journal}{Nat. Mater.} \textbf{\bibinfo{volume}{19}},
  \bibinfo{pages}{522} (\bibinfo{year}{2020}).

\bibitem[{\citenamefont{Lindner et~al.}(2011)\citenamefont{Lindner, Refael, and
  Galitski}}]{lindner2011floquet}
\bibinfo{author}{\bibfnamefont{N.~H.} \bibnamefont{Lindner}},
  \bibinfo{author}{\bibfnamefont{G.}~\bibnamefont{Refael}}, \bibnamefont{and}
  \bibinfo{author}{\bibfnamefont{V.}~\bibnamefont{Galitski}},
  \bibinfo{journal}{Nat. Phys.} \textbf{\bibinfo{volume}{7}},
  \bibinfo{pages}{490} (\bibinfo{year}{2011}).

\bibitem[{\citenamefont{Wang et~al.}(2013)\citenamefont{Wang, Steinberg,
  Jarillo-Herrero, and Gedik}}]{Wang2013Observation}
\bibinfo{author}{\bibfnamefont{Y.~H.} \bibnamefont{Wang}},
  \bibinfo{author}{\bibfnamefont{H.}~\bibnamefont{Steinberg}},
  \bibinfo{author}{\bibfnamefont{P.}~\bibnamefont{Jarillo-Herrero}},
  \bibnamefont{and} \bibinfo{author}{\bibfnamefont{N.}~\bibnamefont{Gedik}},
  \bibinfo{journal}{Science} \textbf{\bibinfo{volume}{342}},
  \bibinfo{pages}{453} (\bibinfo{year}{2013}).

\bibitem[{\citenamefont{McIver et~al.}(2020)\citenamefont{McIver, Schulte,
  Stein, Matsuyama, Jotzu, Meier, and Cavalleri}}]{mciver2020light}
\bibinfo{author}{\bibfnamefont{J.~W.} \bibnamefont{McIver}},
  \bibinfo{author}{\bibfnamefont{B.}~\bibnamefont{Schulte}},
  \bibinfo{author}{\bibfnamefont{F.-U.} \bibnamefont{Stein}},
  \bibinfo{author}{\bibfnamefont{T.}~\bibnamefont{Matsuyama}},
  \bibinfo{author}{\bibfnamefont{G.}~\bibnamefont{Jotzu}},
  \bibinfo{author}{\bibfnamefont{G.}~\bibnamefont{Meier}}, \bibnamefont{and}
  \bibinfo{author}{\bibfnamefont{A.}~\bibnamefont{Cavalleri}},
  \bibinfo{journal}{Nat. Phys.} \textbf{\bibinfo{volume}{16}},
  \bibinfo{pages}{38} (\bibinfo{year}{2020}).

\bibitem[{\citenamefont{Kitagawa et~al.}(2011)\citenamefont{Kitagawa, Oka,
  Brataas, Fu, and Demler}}]{Kitagawa2011Transport}
\bibinfo{author}{\bibfnamefont{T.}~\bibnamefont{Kitagawa}},
  \bibinfo{author}{\bibfnamefont{T.}~\bibnamefont{Oka}},
  \bibinfo{author}{\bibfnamefont{A.}~\bibnamefont{Brataas}},
  \bibinfo{author}{\bibfnamefont{L.}~\bibnamefont{Fu}}, \bibnamefont{and}
  \bibinfo{author}{\bibfnamefont{E.}~\bibnamefont{Demler}},
  \bibinfo{journal}{Phys. Rev. B} \textbf{\bibinfo{volume}{84}},
  \bibinfo{pages}{235108} (\bibinfo{year}{2011}).

\bibitem[{\citenamefont{Xu et~al.}(2021)\citenamefont{Xu, Zhou, and
  Li}}]{xu2021light}
\bibinfo{author}{\bibfnamefont{H.}~\bibnamefont{Xu}},
  \bibinfo{author}{\bibfnamefont{J.}~\bibnamefont{Zhou}}, \bibnamefont{and}
  \bibinfo{author}{\bibfnamefont{J.}~\bibnamefont{Li}}, \bibinfo{journal}{Adv.
  Sci.} \textbf{\bibinfo{volume}{8}}, \bibinfo{pages}{2101508}
  (\bibinfo{year}{2021}).

\bibitem[{\citenamefont{Kong et~al.}(2022)\citenamefont{Kong, Luo, Li, Yoon,
  Berlijn, and Liang}}]{kong2022floquet}
\bibinfo{author}{\bibfnamefont{X.}~\bibnamefont{Kong}},
  \bibinfo{author}{\bibfnamefont{W.}~\bibnamefont{Luo}},
  \bibinfo{author}{\bibfnamefont{L.}~\bibnamefont{Li}},
  \bibinfo{author}{\bibfnamefont{M.}~\bibnamefont{Yoon}},
  \bibinfo{author}{\bibfnamefont{T.}~\bibnamefont{Berlijn}}, \bibnamefont{and}
  \bibinfo{author}{\bibfnamefont{L.}~\bibnamefont{Liang}}, \bibinfo{journal}{2D
  Mater.} \textbf{\bibinfo{volume}{9}}, \bibinfo{pages}{025005}
  (\bibinfo{year}{2022}).

\bibitem[{\citenamefont{Ning et~al.}(2022)\citenamefont{Ning, Zheng, Xu, and
  Wang}}]{ning2022photoinduced}
\bibinfo{author}{\bibfnamefont{Z.}~\bibnamefont{Ning}},
  \bibinfo{author}{\bibfnamefont{B.}~\bibnamefont{Zheng}},
  \bibinfo{author}{\bibfnamefont{D.-H.} \bibnamefont{Xu}}, \bibnamefont{and}
  \bibinfo{author}{\bibfnamefont{R.}~\bibnamefont{Wang}},
  \bibinfo{journal}{Phys. Rev. B} \textbf{\bibinfo{volume}{105}},
  \bibinfo{pages}{035103} (\bibinfo{year}{2022}).

\bibitem[{\citenamefont{Wang et~al.}(2018)\citenamefont{Wang, Liu, Yang, and
  Liu}}]{wang2018Light}
\bibinfo{author}{\bibfnamefont{Z.~F.} \bibnamefont{Wang}},
  \bibinfo{author}{\bibfnamefont{Z.}~\bibnamefont{Liu}},
  \bibinfo{author}{\bibfnamefont{J.}~\bibnamefont{Yang}}, \bibnamefont{and}
  \bibinfo{author}{\bibfnamefont{F.}~\bibnamefont{Liu}},
  \bibinfo{journal}{Phys. Rev. Lett.} \textbf{\bibinfo{volume}{120}},
  \bibinfo{pages}{156406} (\bibinfo{year}{2018}).

\bibitem[{\citenamefont{Yap et~al.}(2017)\citenamefont{Yap, Zhou, Wang, and
  Gong}}]{Yap2017Computational}
\bibinfo{author}{\bibfnamefont{H.~H.} \bibnamefont{Yap}},
  \bibinfo{author}{\bibfnamefont{L.}~\bibnamefont{Zhou}},
  \bibinfo{author}{\bibfnamefont{J.-S.} \bibnamefont{Wang}}, \bibnamefont{and}
  \bibinfo{author}{\bibfnamefont{J.}~\bibnamefont{Gong}},
  \bibinfo{journal}{Phys. Rev. B} \textbf{\bibinfo{volume}{96}},
  \bibinfo{pages}{165443} (\bibinfo{year}{2017}).

\bibitem[{\citenamefont{Liu et~al.}(2021)\citenamefont{Liu, Tang, H{\"u}bener,
  De~Giovannini, Duan, and Rubio}}]{liu2021floquet}
\bibinfo{author}{\bibfnamefont{X.}~\bibnamefont{Liu}},
  \bibinfo{author}{\bibfnamefont{P.}~\bibnamefont{Tang}},
  \bibinfo{author}{\bibfnamefont{H.}~\bibnamefont{H{\"u}bener}},
  \bibinfo{author}{\bibfnamefont{U.}~\bibnamefont{De~Giovannini}},
  \bibinfo{author}{\bibfnamefont{W.}~\bibnamefont{Duan}}, \bibnamefont{and}
  \bibinfo{author}{\bibfnamefont{A.}~\bibnamefont{Rubio}},
  \bibinfo{journal}{arXiv preprint arXiv:2106.06977}  (\bibinfo{year}{2021}).

\bibitem[{\citenamefont{Qin et~al.}(2022)\citenamefont{Qin, Chen, and
  Lu}}]{qin2022phase}
\bibinfo{author}{\bibfnamefont{F.}~\bibnamefont{Qin}},
  \bibinfo{author}{\bibfnamefont{R.}~\bibnamefont{Chen}}, \bibnamefont{and}
  \bibinfo{author}{\bibfnamefont{H.-Z.} \bibnamefont{Lu}}, \bibinfo{journal}{J.
  Phys. Condens. Matter} \textbf{\bibinfo{volume}{34}}, \bibinfo{pages}{225001}
  (\bibinfo{year}{2022}).

\bibitem[{\citenamefont{Yan and Wang}(2016)}]{Yan2016Tunable}
\bibinfo{author}{\bibfnamefont{Z.}~\bibnamefont{Yan}} \bibnamefont{and}
  \bibinfo{author}{\bibfnamefont{Z.}~\bibnamefont{Wang}},
  \bibinfo{journal}{Phys. Rev. Lett.} \textbf{\bibinfo{volume}{117}},
  \bibinfo{pages}{087402} (\bibinfo{year}{2016}).

\bibitem[{\citenamefont{H{\"u}bener et~al.}(2017)\citenamefont{H{\"u}bener,
  Sentef, De~Giovannini, Kemper, and Rubio}}]{hubener2017creating}
\bibinfo{author}{\bibfnamefont{H.}~\bibnamefont{H{\"u}bener}},
  \bibinfo{author}{\bibfnamefont{M.~A.} \bibnamefont{Sentef}},
  \bibinfo{author}{\bibfnamefont{U.}~\bibnamefont{De~Giovannini}},
  \bibinfo{author}{\bibfnamefont{A.~F.} \bibnamefont{Kemper}},
  \bibnamefont{and} \bibinfo{author}{\bibfnamefont{A.}~\bibnamefont{Rubio}},
  \bibinfo{journal}{Nat. Commun.} \textbf{\bibinfo{volume}{8}},
  \bibinfo{pages}{13940} (\bibinfo{year}{2017}).

\bibitem[{\citenamefont{Liu et~al.}(2019)\citenamefont{Liu, Sun, and
  Meng}}]{Liu2019Engineering}
\bibinfo{author}{\bibfnamefont{H.}~\bibnamefont{Liu}},
  \bibinfo{author}{\bibfnamefont{J.-T.} \bibnamefont{Sun}}, \bibnamefont{and}
  \bibinfo{author}{\bibfnamefont{S.}~\bibnamefont{Meng}},
  \bibinfo{journal}{Phys. Rev. B} \textbf{\bibinfo{volume}{99}},
  \bibinfo{pages}{075121} (\bibinfo{year}{2019}).

\bibitem[{\citenamefont{Zhang et~al.}(2018)\citenamefont{Zhang, Wang, Ruan,
  Yao, and Zhang}}]{Zhang2018Engineering}
\bibinfo{author}{\bibfnamefont{D.}~\bibnamefont{Zhang}},
  \bibinfo{author}{\bibfnamefont{H.}~\bibnamefont{Wang}},
  \bibinfo{author}{\bibfnamefont{J.}~\bibnamefont{Ruan}},
  \bibinfo{author}{\bibfnamefont{G.}~\bibnamefont{Yao}}, \bibnamefont{and}
  \bibinfo{author}{\bibfnamefont{H.}~\bibnamefont{Zhang}},
  \bibinfo{journal}{Phys. Rev. B} \textbf{\bibinfo{volume}{97}},
  \bibinfo{pages}{195139} (\bibinfo{year}{2018}).

\bibitem[{\citenamefont{Li et~al.}(2018)\citenamefont{Li, Lee, and
  Gong}}]{Li2018Realistic}
\bibinfo{author}{\bibfnamefont{L.}~\bibnamefont{Li}},
  \bibinfo{author}{\bibfnamefont{C.~H.} \bibnamefont{Lee}}, \bibnamefont{and}
  \bibinfo{author}{\bibfnamefont{J.}~\bibnamefont{Gong}},
  \bibinfo{journal}{Phys. Rev. Lett.} \textbf{\bibinfo{volume}{121}},
  \bibinfo{pages}{036401} (\bibinfo{year}{2018}).

\bibitem[{\citenamefont{Zhu et~al.}(2021{\natexlab{b}})\citenamefont{Zhu, Umer,
  and Gong}}]{Zhu2021Floquet}
\bibinfo{author}{\bibfnamefont{W.}~\bibnamefont{Zhu}},
  \bibinfo{author}{\bibfnamefont{M.}~\bibnamefont{Umer}}, \bibnamefont{and}
  \bibinfo{author}{\bibfnamefont{J.}~\bibnamefont{Gong}},
  \bibinfo{journal}{Phys. Rev. Research} \textbf{\bibinfo{volume}{3}},
  \bibinfo{pages}{L032026} (\bibinfo{year}{2021}{\natexlab{b}}).

\bibitem[{\citenamefont{Zhou et~al.}(2016)\citenamefont{Zhou, Chen, and
  Gong}}]{Zhou2016Floquet}
\bibinfo{author}{\bibfnamefont{L.}~\bibnamefont{Zhou}},
  \bibinfo{author}{\bibfnamefont{C.}~\bibnamefont{Chen}}, \bibnamefont{and}
  \bibinfo{author}{\bibfnamefont{J.}~\bibnamefont{Gong}},
  \bibinfo{journal}{Phys. Rev. B} \textbf{\bibinfo{volume}{94}},
  \bibinfo{pages}{075443} (\bibinfo{year}{2016}).

\bibitem[{\citenamefont{Liu et~al.}(2018)\citenamefont{Liu, Sun, Cheng, Liu,
  and Meng}}]{Liu2018Photoinduced}
\bibinfo{author}{\bibfnamefont{H.}~\bibnamefont{Liu}},
  \bibinfo{author}{\bibfnamefont{J.-T.} \bibnamefont{Sun}},
  \bibinfo{author}{\bibfnamefont{C.}~\bibnamefont{Cheng}},
  \bibinfo{author}{\bibfnamefont{F.}~\bibnamefont{Liu}}, \bibnamefont{and}
  \bibinfo{author}{\bibfnamefont{S.}~\bibnamefont{Meng}},
  \bibinfo{journal}{Phys. Rev. Lett.} \textbf{\bibinfo{volume}{120}},
  \bibinfo{pages}{237403} (\bibinfo{year}{2018}).

\bibitem[{\citenamefont{G\'omez-Le\'on and
  Platero}(2013)}]{Gomez-Leon2013Floquet}
\bibinfo{author}{\bibfnamefont{A.}~\bibnamefont{G\'omez-Le\'on}}
  \bibnamefont{and} \bibinfo{author}{\bibfnamefont{G.}~\bibnamefont{Platero}},
  \bibinfo{journal}{Phys. Rev. Lett.} \textbf{\bibinfo{volume}{110}},
  \bibinfo{pages}{200403} (\bibinfo{year}{2013}).

\bibitem[{\citenamefont{Rudner et~al.}(2013)\citenamefont{Rudner, Lindner,
  Berg, and Levin}}]{Rudner2013Anomalous}
\bibinfo{author}{\bibfnamefont{M.~S.} \bibnamefont{Rudner}},
  \bibinfo{author}{\bibfnamefont{N.~H.} \bibnamefont{Lindner}},
  \bibinfo{author}{\bibfnamefont{E.}~\bibnamefont{Berg}}, \bibnamefont{and}
  \bibinfo{author}{\bibfnamefont{M.}~\bibnamefont{Levin}},
  \bibinfo{journal}{Phys. Rev. X} \textbf{\bibinfo{volume}{3}},
  \bibinfo{pages}{031005} (\bibinfo{year}{2013}).

\bibitem[{\citenamefont{Ezawa}(2013)}]{Ezawa2013Photoinduced}
\bibinfo{author}{\bibfnamefont{M.}~\bibnamefont{Ezawa}},
  \bibinfo{journal}{Phys. Rev. Lett.} \textbf{\bibinfo{volume}{110}},
  \bibinfo{pages}{026603} (\bibinfo{year}{2013}).

\bibitem[{\citenamefont{Grushin et~al.}(2014)\citenamefont{Grushin,
  G\'omez-Le\'on, and Neupert}}]{Grushin2014Floquet}
\bibinfo{author}{\bibfnamefont{A.~G.} \bibnamefont{Grushin}},
  \bibinfo{author}{\bibfnamefont{A.}~\bibnamefont{G\'omez-Le\'on}},
  \bibnamefont{and} \bibinfo{author}{\bibfnamefont{T.}~\bibnamefont{Neupert}},
  \bibinfo{journal}{Phys. Rev. Lett.} \textbf{\bibinfo{volume}{112}},
  \bibinfo{pages}{156801} (\bibinfo{year}{2014}).

\bibitem[{\citenamefont{Mahmood et~al.}(2016)\citenamefont{Mahmood, Chan,
  Alpichshev, Gardner, Lee, Lee, and Gedik}}]{mahmood2016selective}
\bibinfo{author}{\bibfnamefont{F.}~\bibnamefont{Mahmood}},
  \bibinfo{author}{\bibfnamefont{C.-K.} \bibnamefont{Chan}},
  \bibinfo{author}{\bibfnamefont{Z.}~\bibnamefont{Alpichshev}},
  \bibinfo{author}{\bibfnamefont{D.}~\bibnamefont{Gardner}},
  \bibinfo{author}{\bibfnamefont{Y.}~\bibnamefont{Lee}},
  \bibinfo{author}{\bibfnamefont{P.~A.} \bibnamefont{Lee}}, \bibnamefont{and}
  \bibinfo{author}{\bibfnamefont{N.}~\bibnamefont{Gedik}},
  \bibinfo{journal}{Nat. Phys.} \textbf{\bibinfo{volume}{12}},
  \bibinfo{pages}{306} (\bibinfo{year}{2016}).

\bibitem[{\citenamefont{Oka and Kitamura}(2019)}]{Oka2019Floquet}
\bibinfo{author}{\bibfnamefont{T.}~\bibnamefont{Oka}} \bibnamefont{and}
  \bibinfo{author}{\bibfnamefont{S.}~\bibnamefont{Kitamura}},
  \bibinfo{journal}{Annu. Rev. Condens. Matter Phys.}
  \textbf{\bibinfo{volume}{10}}, \bibinfo{pages}{387} (\bibinfo{year}{2019}).

\bibitem[{\citenamefont{Mikami et~al.}(2016)\citenamefont{Mikami, Kitamura,
  Yasuda, Tsuji, Oka, and Aoki}}]{Mikami2016Brillouin}
\bibinfo{author}{\bibfnamefont{T.}~\bibnamefont{Mikami}},
  \bibinfo{author}{\bibfnamefont{S.}~\bibnamefont{Kitamura}},
  \bibinfo{author}{\bibfnamefont{K.}~\bibnamefont{Yasuda}},
  \bibinfo{author}{\bibfnamefont{N.}~\bibnamefont{Tsuji}},
  \bibinfo{author}{\bibfnamefont{T.}~\bibnamefont{Oka}}, \bibnamefont{and}
  \bibinfo{author}{\bibfnamefont{H.}~\bibnamefont{Aoki}},
  \bibinfo{journal}{Phys. Rev. B} \textbf{\bibinfo{volume}{93}},
  \bibinfo{pages}{144307} (\bibinfo{year}{2016}).

\bibitem[{\citenamefont{Eckardt and Anisimovas}(2015)}]{Eckardt2015High}
\bibinfo{author}{\bibfnamefont{A.}~\bibnamefont{Eckardt}} \bibnamefont{and}
  \bibinfo{author}{\bibfnamefont{E.}~\bibnamefont{Anisimovas}},
  \bibinfo{journal}{New J. Phys.} \textbf{\bibinfo{volume}{17}},
  \bibinfo{pages}{093039} (\bibinfo{year}{2015}).

\bibitem[{\citenamefont{Ma et~al.}(2021)\citenamefont{Ma, Sun, Liu, and
  Zhao}}]{Ma2021Floquet}
\bibinfo{author}{\bibfnamefont{X.}~\bibnamefont{Ma}},
  \bibinfo{author}{\bibfnamefont{L.}~\bibnamefont{Sun}},
  \bibinfo{author}{\bibfnamefont{J.}~\bibnamefont{Liu}}, \bibnamefont{and}
  \bibinfo{author}{\bibfnamefont{M.}~\bibnamefont{Zhao}},
  \bibinfo{journal}{Phys. Rev. B} \textbf{\bibinfo{volume}{104}},
  \bibinfo{pages}{155439} (\bibinfo{year}{2021}).

\bibitem[{\citenamefont{Pervishko et~al.}(2018)\citenamefont{Pervishko, Yudin,
  and Shelykh}}]{Pervishko2018Impact}
\bibinfo{author}{\bibfnamefont{A.~A.} \bibnamefont{Pervishko}},
  \bibinfo{author}{\bibfnamefont{D.}~\bibnamefont{Yudin}}, \bibnamefont{and}
  \bibinfo{author}{\bibfnamefont{I.~A.} \bibnamefont{Shelykh}},
  \bibinfo{journal}{Phys. Rev. B} \textbf{\bibinfo{volume}{97}},
  \bibinfo{pages}{075420} (\bibinfo{year}{2018}).

\bibitem[{\citenamefont{Cheng et~al.}(2019)\citenamefont{Cheng, Pan, Wang,
  Zhang, Yu, Gover, Zhang, Li, Zhou, and Zhu}}]{Cheng2019}
\bibinfo{author}{\bibfnamefont{Q.}~\bibnamefont{Cheng}},
  \bibinfo{author}{\bibfnamefont{Y.}~\bibnamefont{Pan}},
  \bibinfo{author}{\bibfnamefont{H.}~\bibnamefont{Wang}},
  \bibinfo{author}{\bibfnamefont{C.}~\bibnamefont{Zhang}},
  \bibinfo{author}{\bibfnamefont{D.}~\bibnamefont{Yu}},
  \bibinfo{author}{\bibfnamefont{A.}~\bibnamefont{Gover}},
  \bibinfo{author}{\bibfnamefont{H.}~\bibnamefont{Zhang}},
  \bibinfo{author}{\bibfnamefont{T.}~\bibnamefont{Li}},
  \bibinfo{author}{\bibfnamefont{L.}~\bibnamefont{Zhou}}, \bibnamefont{and}
  \bibinfo{author}{\bibfnamefont{S.}~\bibnamefont{Zhu}},
  \bibinfo{journal}{Phys. Rev. Lett.} \textbf{\bibinfo{volume}{122}},
  \bibinfo{pages}{173901} (\bibinfo{year}{2019}).

\bibitem[{\citenamefont{Wang et~al.}(2017)\citenamefont{Wang, Chen, Bomantara,
  Gong, and Xing}}]{Wang2017}
\bibinfo{author}{\bibfnamefont{H.-Q.} \bibnamefont{Wang}},
  \bibinfo{author}{\bibfnamefont{M.~N.} \bibnamefont{Chen}},
  \bibinfo{author}{\bibfnamefont{R.~W.} \bibnamefont{Bomantara}},
  \bibinfo{author}{\bibfnamefont{J.}~\bibnamefont{Gong}}, \bibnamefont{and}
  \bibinfo{author}{\bibfnamefont{D.~Y.} \bibnamefont{Xing}},
  \bibinfo{journal}{Phys. Rev. B} \textbf{\bibinfo{volume}{95}},
  \bibinfo{pages}{075136} (\bibinfo{year}{2017}).

\bibitem[{\citenamefont{Bomantara et~al.}(2016)\citenamefont{Bomantara,
  Raghava, Zhou, and Gong}}]{Bomantara2016}
\bibinfo{author}{\bibfnamefont{R.~W.} \bibnamefont{Bomantara}},
  \bibinfo{author}{\bibfnamefont{G.~N.} \bibnamefont{Raghava}},
  \bibinfo{author}{\bibfnamefont{L.}~\bibnamefont{Zhou}}, \bibnamefont{and}
  \bibinfo{author}{\bibfnamefont{J.}~\bibnamefont{Gong}},
  \bibinfo{journal}{Phys. Rev. E} \textbf{\bibinfo{volume}{93}},
  \bibinfo{pages}{022209} (\bibinfo{year}{2016}).

\bibitem[{\citenamefont{Bomantara and Gong}(2018)}]{Bomantara2018}
\bibinfo{author}{\bibfnamefont{R.~W.} \bibnamefont{Bomantara}}
  \bibnamefont{and} \bibinfo{author}{\bibfnamefont{J.}~\bibnamefont{Gong}},
  \bibinfo{journal}{Phys. Rev. Lett.} \textbf{\bibinfo{volume}{120}},
  \bibinfo{pages}{230405} (\bibinfo{year}{2018}).

\bibitem[{\citenamefont{Eremeev et~al.}(2012)\citenamefont{Eremeev, Vergniory,
  Menshchikova, Shaposhnikov, and Chulkov}}]{Eremeev2012effect}
\bibinfo{author}{\bibfnamefont{S.~V.} \bibnamefont{Eremeev}},
  \bibinfo{author}{\bibfnamefont{M.~G.} \bibnamefont{Vergniory}},
  \bibinfo{author}{\bibfnamefont{T.~V.} \bibnamefont{Menshchikova}},
  \bibinfo{author}{\bibfnamefont{A.~A.} \bibnamefont{Shaposhnikov}},
  \bibnamefont{and} \bibinfo{author}{\bibfnamefont{E.~V.}
  \bibnamefont{Chulkov}}, \bibinfo{journal}{New J. Phys.}
  \textbf{\bibinfo{volume}{14}}, \bibinfo{pages}{113030}
  (\bibinfo{year}{2012}).

\bibitem[{\citenamefont{Wang et~al.}(2022)\citenamefont{Wang, Wang, Xing, and
  Zhang}}]{wang2022three}
\bibinfo{author}{\bibfnamefont{D.}~\bibnamefont{Wang}},
  \bibinfo{author}{\bibfnamefont{H.}~\bibnamefont{Wang}},
  \bibinfo{author}{\bibfnamefont{D.}~\bibnamefont{Xing}}, \bibnamefont{and}
  \bibinfo{author}{\bibfnamefont{H.}~\bibnamefont{Zhang}},
  \bibinfo{journal}{arXiv:2205.08204}  (\bibinfo{year}{2022}).

\bibitem[{\citenamefont{Kresse and Furthm\"uller}(1996)}]{Kresse1996vasp}
\bibinfo{author}{\bibfnamefont{G.}~\bibnamefont{Kresse}} \bibnamefont{and}
  \bibinfo{author}{\bibfnamefont{J.}~\bibnamefont{Furthm\"uller}},
  \bibinfo{journal}{Phys. Rev. B} \textbf{\bibinfo{volume}{54}},
  \bibinfo{pages}{11169} (\bibinfo{year}{1996}).

\bibitem[{\citenamefont{Kresse and Joubert}(1999)}]{Kresse1999vasp}
\bibinfo{author}{\bibfnamefont{G.}~\bibnamefont{Kresse}} \bibnamefont{and}
  \bibinfo{author}{\bibfnamefont{D.}~\bibnamefont{Joubert}},
  \bibinfo{journal}{Phys. Rev. B} \textbf{\bibinfo{volume}{59}},
  \bibinfo{pages}{1758} (\bibinfo{year}{1999}).

\bibitem[{\citenamefont{Perdew et~al.}(1996)\citenamefont{Perdew, Burke, and
  Ernzerhof}}]{Perdew1996PBE}
\bibinfo{author}{\bibfnamefont{J.~P.} \bibnamefont{Perdew}},
  \bibinfo{author}{\bibfnamefont{K.}~\bibnamefont{Burke}}, \bibnamefont{and}
  \bibinfo{author}{\bibfnamefont{M.}~\bibnamefont{Ernzerhof}},
  \bibinfo{journal}{Phys. Rev. Lett.} \textbf{\bibinfo{volume}{77}},
  \bibinfo{pages}{3865} (\bibinfo{year}{1996}).

\bibitem[{\citenamefont{Bl\"ochl}(1994)}]{Blochl1994PBE}
\bibinfo{author}{\bibfnamefont{P.~E.} \bibnamefont{Bl\"ochl}},
  \bibinfo{journal}{Phys. Rev. B} \textbf{\bibinfo{volume}{50}},
  \bibinfo{pages}{17953} (\bibinfo{year}{1994}).

\bibitem[{\citenamefont{Marzari and Vanderbilt}(1997)}]{marzari1997maximally}
\bibinfo{author}{\bibfnamefont{N.}~\bibnamefont{Marzari}} \bibnamefont{and}
  \bibinfo{author}{\bibfnamefont{D.}~\bibnamefont{Vanderbilt}},
  \bibinfo{journal}{Phys. Rev. B} \textbf{\bibinfo{volume}{56}},
  \bibinfo{pages}{12847} (\bibinfo{year}{1997}).

\bibitem[{\citenamefont{Souza et~al.}(2001)\citenamefont{Souza, Marzari, and
  Vanderbilt}}]{souza2001maximally}
\bibinfo{author}{\bibfnamefont{I.}~\bibnamefont{Souza}},
  \bibinfo{author}{\bibfnamefont{N.}~\bibnamefont{Marzari}}, \bibnamefont{and}
  \bibinfo{author}{\bibfnamefont{D.}~\bibnamefont{Vanderbilt}},
  \bibinfo{journal}{Phys. Rev. B} \textbf{\bibinfo{volume}{65}},
  \bibinfo{pages}{035109} (\bibinfo{year}{2001}).

\bibitem[{\citenamefont{Pizzi et~al.}(2020)\citenamefont{Pizzi, Vitale, Arita,
  Blügel, Freimuth, G{\'{e}}ranton, Gibertini, Gresch, Johnson, Koretsune
  et~al.}}]{Pizzi2020wannier90}
\bibinfo{author}{\bibfnamefont{G.}~\bibnamefont{Pizzi}},
  \bibinfo{author}{\bibfnamefont{V.}~\bibnamefont{Vitale}},
  \bibinfo{author}{\bibfnamefont{R.}~\bibnamefont{Arita}},
  \bibinfo{author}{\bibfnamefont{S.}~\bibnamefont{Blügel}},
  \bibinfo{author}{\bibfnamefont{F.}~\bibnamefont{Freimuth}},
  \bibinfo{author}{\bibfnamefont{G.}~\bibnamefont{G{\'{e}}ranton}},
  \bibinfo{author}{\bibfnamefont{M.}~\bibnamefont{Gibertini}},
  \bibinfo{author}{\bibfnamefont{D.}~\bibnamefont{Gresch}},
  \bibinfo{author}{\bibfnamefont{C.}~\bibnamefont{Johnson}},
  \bibinfo{author}{\bibfnamefont{T.}~\bibnamefont{Koretsune}},
  \bibnamefont{et~al.}, \bibinfo{journal}{J. Phys. Condens. Matter}
  \textbf{\bibinfo{volume}{32}}, \bibinfo{pages}{165902}
  (\bibinfo{year}{2020}).

\bibitem[{\citenamefont{Sancho et~al.}(1984)\citenamefont{Sancho, Sancho, and
  Rubio}}]{Sancho1984Quick}
\bibinfo{author}{\bibfnamefont{M.~P.~L.} \bibnamefont{Sancho}},
  \bibinfo{author}{\bibfnamefont{J.~M.~L.} \bibnamefont{Sancho}},
  \bibnamefont{and} \bibinfo{author}{\bibfnamefont{J.}~\bibnamefont{Rubio}},
  \bibinfo{journal}{J. Phys. F: Met. Phys} \textbf{\bibinfo{volume}{14}},
  \bibinfo{pages}{1205} (\bibinfo{year}{1984}).

\bibitem[{\citenamefont{Sancho et~al.}(1985)\citenamefont{Sancho, Sancho,
  Sancho, and Rubio}}]{Sancho1985Highly}
\bibinfo{author}{\bibfnamefont{M.~P.~L.} \bibnamefont{Sancho}},
  \bibinfo{author}{\bibfnamefont{J.~M.~L.} \bibnamefont{Sancho}},
  \bibinfo{author}{\bibfnamefont{J.~M.~L.} \bibnamefont{Sancho}},
  \bibnamefont{and} \bibinfo{author}{\bibfnamefont{J.}~\bibnamefont{Rubio}},
  \bibinfo{journal}{J. Phys. F: Met. Phys} \textbf{\bibinfo{volume}{15}},
  \bibinfo{pages}{851} (\bibinfo{year}{1985}).

\bibitem[{\citenamefont{He}(2020)}]{he2020mnbi}
\bibinfo{author}{\bibfnamefont{K.}~\bibnamefont{He}}, \bibinfo{journal}{npj
  Quantum Mater.} \textbf{\bibinfo{volume}{5}}, \bibinfo{pages}{90}
  (\bibinfo{year}{2020}).

\bibitem[{\citenamefont{Zhao et~al.}(2020)\citenamefont{Zhao, Zhang, Mei, Zhou,
  Yi, Zhang, Yu, Xiao, Wang, Samarth et~al.}}]{zhao2020tuning}
\bibinfo{author}{\bibfnamefont{Y.-F.} \bibnamefont{Zhao}},
  \bibinfo{author}{\bibfnamefont{R.}~\bibnamefont{Zhang}},
  \bibinfo{author}{\bibfnamefont{R.}~\bibnamefont{Mei}},
  \bibinfo{author}{\bibfnamefont{L.-J.} \bibnamefont{Zhou}},
  \bibinfo{author}{\bibfnamefont{H.}~\bibnamefont{Yi}},
  \bibinfo{author}{\bibfnamefont{Y.-Q.} \bibnamefont{Zhang}},
  \bibinfo{author}{\bibfnamefont{J.}~\bibnamefont{Yu}},
  \bibinfo{author}{\bibfnamefont{R.}~\bibnamefont{Xiao}},
  \bibinfo{author}{\bibfnamefont{K.}~\bibnamefont{Wang}},
  \bibinfo{author}{\bibfnamefont{N.}~\bibnamefont{Samarth}},
  \bibnamefont{et~al.}, \bibinfo{journal}{Nature}
  \textbf{\bibinfo{volume}{588}}, \bibinfo{pages}{419} (\bibinfo{year}{2020}).

\bibitem[{\citenamefont{Zhao et~al.}(2022)\citenamefont{Zhao, Zhang, Zhou, Mei,
  Yan, Chan, Liu, and Chang}}]{Zhao2022Zero}
\bibinfo{author}{\bibfnamefont{Y.-F.} \bibnamefont{Zhao}},
  \bibinfo{author}{\bibfnamefont{R.}~\bibnamefont{Zhang}},
  \bibinfo{author}{\bibfnamefont{L.-J.} \bibnamefont{Zhou}},
  \bibinfo{author}{\bibfnamefont{R.}~\bibnamefont{Mei}},
  \bibinfo{author}{\bibfnamefont{Z.-J.} \bibnamefont{Yan}},
  \bibinfo{author}{\bibfnamefont{M.~H.~W.} \bibnamefont{Chan}},
  \bibinfo{author}{\bibfnamefont{C.-X.} \bibnamefont{Liu}}, \bibnamefont{and}
  \bibinfo{author}{\bibfnamefont{C.-Z.} \bibnamefont{Chang}},
  \bibinfo{journal}{Phys. Rev. Lett.} \textbf{\bibinfo{volume}{128}},
  \bibinfo{pages}{216801} (\bibinfo{year}{2022}).

\bibitem[{\citenamefont{Ge et~al.}(2020)\citenamefont{Ge, Liu, Li, Li, Luo, Wu,
  Xu, and Wang}}]{ge2020high}
\bibinfo{author}{\bibfnamefont{J.}~\bibnamefont{Ge}},
  \bibinfo{author}{\bibfnamefont{Y.}~\bibnamefont{Liu}},
  \bibinfo{author}{\bibfnamefont{J.}~\bibnamefont{Li}},
  \bibinfo{author}{\bibfnamefont{H.}~\bibnamefont{Li}},
  \bibinfo{author}{\bibfnamefont{T.}~\bibnamefont{Luo}},
  \bibinfo{author}{\bibfnamefont{Y.}~\bibnamefont{Wu}},
  \bibinfo{author}{\bibfnamefont{Y.}~\bibnamefont{Xu}}, \bibnamefont{and}
  \bibinfo{author}{\bibfnamefont{J.}~\bibnamefont{Wang}},
  \bibinfo{journal}{Natl. Sci. Rev.} \textbf{\bibinfo{volume}{7}},
  \bibinfo{pages}{1280} (\bibinfo{year}{2020}).

\end{thebibliography}
\end{document}